\documentclass[aps, prl, superscriptaddress, showpacs, twocolumn, reprint]{revtex4-1}
\usepackage{units}
\usepackage{amsmath}
\usepackage{amssymb}
\usepackage{graphicx}
\usepackage{bm}
\usepackage{multirow,color,relsize,ulem,microtype,mathtools}
\usepackage{natbib}

\newcommand{\be}{\begin{equation}}
\newcommand{\ee}{\end{equation}}

\newcommand{\wn}{\omega_0}

\begin{document}
\title{Perfectly absorbing exceptional points and chiral absorbers}

\author{William R. Sweeney}
\email{william.sweeney@yale.edu}
\affiliation{Department of Physics, Yale University, New Haven, CT 06520, USA}
\author{Chia Wei Hsu}
\affiliation{Department of Applied Physics, Yale University, New Haven, CT 06520, USA}
\author{Stefan Rotter}
\affiliation{Institute for Theoretical Physics, Vienna University of Technology (TU Wien), A-1040 Vienna, Austria, EU}
\author{A. Douglas Stone}
\affiliation{Department of Applied Physics, Yale University, New Haven, CT 06520, USA}
\affiliation{Yale Quantum Institute, Yale University, New Haven, CT 06520, USA}
\date{\today}

\begin{abstract}
We identify a new kind of physically realizable exceptional point (EP) corresponding to degenerate coherent perfect absorption, in which two purely incoming solutions of the wave operator for electromagnetic or acoustic waves coalesce to a single state.
Such non-hermitian degeneracies can occur at a real-valued frequency without any associated noise or non-linearity, in contrast to EPs in lasers.
The absorption lineshape for the eigenchannel near the EP is quartic in frequency around its maximum in any dimension.
In general, for the parameters at which an operator EP occurs, the associated scattering matrix does not have an EP.
However, in one dimension, when the $S$-matrix does have a perfectly absorbing EP, it takes on a universal one-parameter form with degenerate values for all scattering coefficients.
For absorbing disk resonators, these EPs give rise to chiral absorption: perfect absorption for only one sense of rotation of the input wave.
\end{abstract}

\pacs{}

\maketitle


Exceptional points (EPs) are generic degeneracies of non-hermitian systems, where two eigenvalues {\it and} eigenvectors of a linear operator coalesce, reducing the size of the space spanned by the eigenbasis~\cite{Kato:1995bm, Moiseyev:2011tx, Heiss:2012vja, Feng:2017ww, ElGanainy:2018ie}.
EPs arise in open physical systems and are of interest for a number of reasons.
For example, they induce chiral behavior under cyclic variation of the parameters of the relevant operator, leading to robust asymmetric state transfer~\cite{Doppler:2016ke, Xu:ko}.
In addition, near an EP a resonant system shows enhanced frequency splitting under small perturbations that may lead to improved sensing~\cite{Chen:2017jk, Hodaei:fe, Zhang:2018tq}.
EPs can lead to counter-intuitive behavior as loss or gain is varied, such as resonance trapping in nuclear and atomic scattering~\cite{Persson:2000wo, Rotter:2009fr}, enhanced transmission with increasing loss in coupled waveguides~\cite{Guo:2009hd, Ruter:2010dy}, and suppression of lasing with increasing gain in coupled cavity systems~\cite{Liertzer:2012hl, Brandstetter:2014bt}.
Recently, work of Wiersig has shown that the chirality associated with EPs can be manifested in disk resonators in the form of chiral lasing~\cite{Wiersig:2011hs, Wiersig:2014bq}, an effect confirmed in recent experiments by Peng, et al.~\cite{Peng:2016dd}.

Two types of EPs have been extensively studied in physics: resonant and scattering.
First to be studied were resonant EPs, in which two resonances of an open system coalesce.
Resonances are solutions of the wave equation with purely outgoing boundary conditions, typically occurring at complex-valued frequencies, corresponding to poles of the scattering matrix $S$.
When parameters in the wave equation are varied, it is possible for two such resonances to coincide (double pole), leading generically to an EP.
In unitary systems (e.g.~no imaginary part of the index of refraction or potential), resonant EPs can only occur at complex frequencies (energies) below the real axis, and do not correspond to physical steady-state solutions, although they can still strongly influence the scattering properties for real frequencies~\cite{Peng:2014kl, Zhen:bl, Zhou:2018dy}.
By adding gain to an electromagnetic cavity one may bring the resonant EP to a real frequency, corresponding to lasing at threshold.
But an amplifying system is not ideal for the study of EPs, due to the large amplified spontaneous emission noise at threshold, and the necessity of including the non-linearity of the medium to stabilize lasing above threshold.

Scattering EPs are EPs of $S$ and have mainly been studied in systems with balanced loss and gain ($\mathcal{PT}$ symmetry and related variants), where the scattering eigenchannels make a transition from flux-conserving to amplifying or attenuating propagation~\cite{Chong:2011ev, Ge:2012bq, Ambichl:2013gq}.
Typical eigenstates of $S$ have both incoming and outgoing components, and hence are not resonances of the system.

Here we study a new kind of EP, the coalescence of two solutions of the wave operator with purely {\it incoming} boundary conditions, corresponding to perfect absorption.
When a {\it single} such wave solution occurs at a real frequency, it is an example of Coherent Perfect Absorption (CPA)~\cite{Chong:2010ft, Wan:2011bz, Noh:2012wx, Piper:2014js, Zhou:2016tv, Zhao:2016cd, Baranov:2017jv}, a variant and generalization of the concept of critical coupling~\cite{Yariv:2000dz}, in which a particular steady-state incident wavefront is completely absorbed.  
The specific input state is the time-reverse of the threshold lasing mode for the same cavity, but with gain replacing loss [$n(\vec{r}) \rightarrow n^*(\vec{r})$].
Achieving CPA typically requires tuning the input frequency and the degree of absorption.
With no gain or loss, the frequencies of purely incoming/outgoing states occur in conjugate frequency pairs, $\omega_n \pm i \gamma_n$; the addition of material loss is necessary to move the frequency of a purely incoming state onto the real axis to achieve CPA.  Here we study a CPA EP, where two incoming solutions of the wave equation coalesce at a real frequency.
The degeneracy of two eigenfrequencies of the incoming wave operator is generically a CPA EP. Exceptions occur for degenerate but decoupled states, e.g.~those with different symmetry~\cite{Piper:2014js}; these cases will be neglected here.
Such absorbing EPs have not been studied before, but should be readily observable with set-ups previously used to investigate resonant EPs~\cite{Peng:2014kl, Chang:2014by, Miao:2016eo}.

The signature of CPA EP in scattering is a quartic behavior (flattening) of the absorption lineshape in the perfectly absorbed channel (see Fig.~1a-f); for ordinary CPA it is quadratic.
The perfectly absorbed input channel corresponds to an eigenvector of $S$ with eigenvalue zero. To our knowledge, any modification of a lineshape associated with an EP has not been previously observed.
The quartic behavior generalizes to higher dimensional and/or multichannel, quasi-1D CPA EPs as well; but {\it only} in the CPA eigenchannel, and not in the individual scattering coefficients or other eigenchannels (see Fig.~1d).
Its origin can be understood as follows: near an ordinary CPA frequency $\wn$, an eigenvalue of $S$, $\sigma(\omega)$, will pass through zero linearly in the deviation $\delta\equiv\omega-\wn$, so that $|\sigma(\omega)|^2\propto\delta^2$.
In the vicinity of the parameter values leading to CPA EP, there are two CPA frequencies near each other ($\wn + \delta_1$ and $\wn + \delta_2$), both belonging to the same eigenvalue $\sigma(\omega)$, whose smooth variation implies $\sigma(\omega) \propto \delta_1  \delta_2$.
At CPA EP, $\delta_1\rightarrow\delta_2\equiv\delta$, and $|\sigma|^2\propto \delta^4$, which is the quartic absorption lineshape.
The other conceivable behavior, where distinct $S$-matrix eigenvalues meet at zero, does not correspond to CPA EP, but rather to an EP of $S$; the smoothness assumption used above is violated and the lineshape is not quartic.

The general properties described above are exemplified by a one-dimensional electromagnetic structure, consisting of two cavities created by a series of three mirrors (see Fig.~1a-i).
An EP is realized by coupling the two cavities via a central partially reflecting Bragg mirror and introducing unequal absorption within each cavity.
We show three interesting cases.
In Fig.~1a-c, the structure is terminated on the right by a perfect mirror and is accessible only from the left through a partial Bragg mirror, so that $S$ is a scalar, namely, the left reflection amplitude $r_L$.
The absorption is $1- |r_L|^2$.
This set-up corresponds to the usual critical coupling to a cavity (one-channel CPA), except that the cavity is tuned to an EP of the incoming wave operator and hence the absorption lineshape is quartic.
On the other hand, in Fig.~1d-f, the Bragg mirrors on the two ends are both permeable and define a two-channel $S$-matrix, characterized by three scattering amplitudes $r_L$, $r_R$, $t$.
Here, exciting the absorbing eigenchannel of $S$ requires coherent illumination from both sides with a definite relative intensity and phase \cite{Chong:2010ft}.
As shown in Fig.~1d, the quartic absorption lineshape is evident for this input state; however neither the one-sided scattering coefficients ($|r_L|^2$, $|r_R|^2$, $|t|^2$), nor the non-zero eigenchannel exhibit such a flat-top profile.   

\begin{figure}[t]
\centering
\includegraphics[width=\linewidth]{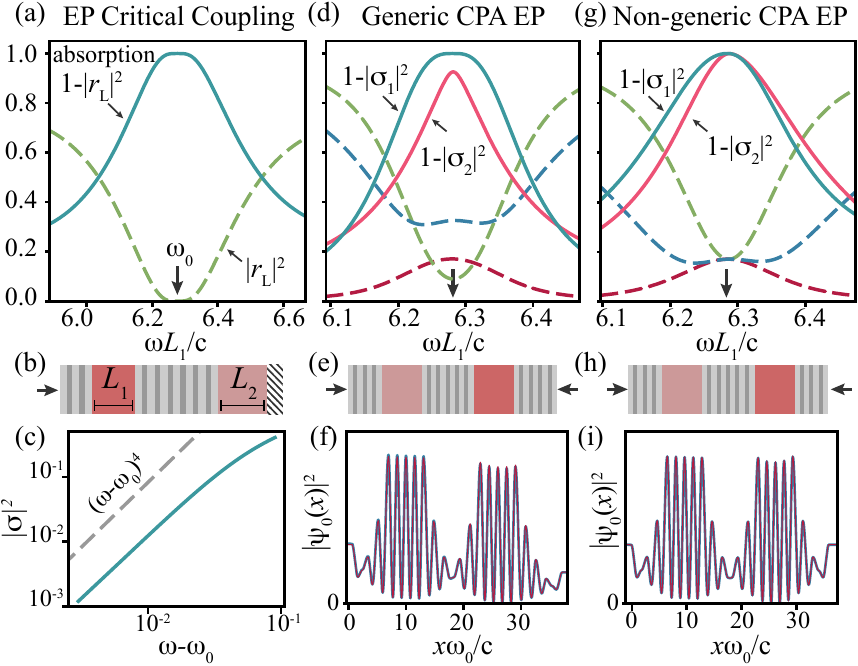}
\caption{
(Color online) Scattering from coupled cavity structures at CPA EP for asymmetric (a-f) and symmetric (g-i) end mirrors.
\textbf{(a,d,g)} Absorption lineshapes of eigenchannels (solid): CPA channel (blue) reaches 100\% absorption at the EP frequency $\omega_0$.
In (a,d) CPA lineshape is quartic [see {\bf(c)}], while in (d) non-CPA channel (red) is quadratic.
In (g) we have an EP of $S$ and the lineshape isn't quartic.
Scattering coefflcients $|r_L|^2$, $|r_R|^2$, $|t|^2$ are shown as green, blue, red dashed lines; in (g) they become degenerate at $\omega_0$, as predicted for an EP of $S$-matrix at zero.
\textbf{(b,e,h)} Schematics of structures: cavities (red) with lengths $L_1, L_2$, and unequal absorption, emitting to 
free space through end mirrors. Right mirror is perfect in (b), permeable but unequal to left in (e) and equal to left in (h); parameter values are given in supplement S1.
\textbf{(f,i)} CPA EP modes: generic case of unequal coupling (f) yields asymmetric asymptotic values for $|\psi|^2$, implying $S$ not at EP. The CPA mode for non-generic case of equal coupling (i) has equal asymptotic values, implying that $S$ is at an EP.
}
\end{figure}

While Fig.~1a-f describe the generic scattering behavior near a CPA EP, there is a novel and interesting non-generic case, exemplified by Fig.~1g-i, which can be realized in the same type of geometry, and does {\it not} show the generic quartic lineshape, but has different and striking scattering properties.
This is a case where CPA EP and an EP of the $S$-matrix approximately coincide.
Hence we now discuss the relationship between exceptional behavior of the wave operator and of $S$.

Every eigenstate of the wave operator with incoming boundary conditions also corresponds to an eigenvector of $S$ with eigenvalue zero. However, the coalescence of two incoming states does {\it not} simultaneously generate an EP of $S$, 
as we now prove.

For simplicity, consider an arbitrary one-dimensional cavity described by the Helmholtz equation:
\begin{equation} \label{Helmholtz}
\{\nabla^2 + \varepsilon (x) k_j^2 \} \psi_j(x) = 0,
\end{equation}
where $\varepsilon (x)$ is the dielectric function of the medium, $k_j = \omega_j/c$, and $\omega_j$ are the discrete complex eigenfrequencies with purely incoming boundary conditions.
Consider two eigenfrequencies, $\omega_1,\omega_2$, initially with different values and linearly independent solutions, $\psi_1 (x),\psi_2 (x)$.
Further assume that tuning $\varepsilon (x)$ causes these two solutions to coalesce at $\omega_0$: $\psi_1,\psi_2 \rightarrow \psi_0$.
By using the wave equation \eqref{Helmholtz} and taking the limit $\omega_1 \to \omega_2 \equiv \wn$, one can derive the identity (see supplement S2)
\begin{equation} \label{bulk integral}
-2i \omega_0 \int_{\mathrm{cav}} dx\, \psi_0 (x) \varepsilon (x) \psi_0 (x) = c_0\, \hat{s}_0 \cdot \hat{s}_0,
\end{equation}
where $\hat{s}_0$ is the normalized eigenvector of the $S$-matrix corresponding to $\psi_0$ (i.e. with eigenvalue zero), and $c_0$ is a 
system-specific constant. 
The integral on the left hand side of Eq.~\eqref{bulk integral} in general does not vanish.
Solutions of the wave equation with either purely incoming {\it or} outgoing boundary conditions do not satisfy any simple biorthogonality relation over the scattering region~\cite{Leung:1994fq, Moiseyev:2011tx}.
Hence at an EP of the incoming wave operator, integrals of this type are non-zero (no self-orthogonality of the EP eigenfunction).
On the other hand, the RHS of Eq.~\eqref{bulk integral} is proportional to the biorthogonal norm of the eigenvector of the symmetric $S$-matrix with eigenvalue zero; as such it vanishes iff $S$ is {\it also} at an EP~\cite{Trefethen:2005wt}.
A non-vanishing LHS implies that CPA EP does not in general correspond to an EP of $S$; indeed for the generic case shown in Fig.~1d-f the $S$-matrix has a second eigenvector which is not perfectly absorbed at CPA EP (red solid line), and hence has non-zero scattering.
This proof generalizes to higher dimensional scattering geometries using Green's theorem. 

Conversely, one can find scattering geometries and structures for which an EP of $S$ can occur for eigenvalue equal to zero; however this does {\it not} in general imply CPA EP.
The EP of $S$ at zero is a specific case of a scattering EP of the type mentioned above{~\cite{Chong:2011ev, Ge:2012bq, Ambichl:2013gq}; we discuss its implications briefly below.
The general case of scattering EPs will be discussed elsewhere~\cite{Sweeney:2018nm}.

A $2\times2$ $S$-matrix with zero eigenvalue, tuned to an EP at frequency $\wn$, satisfies $r_L(\wn)= - r_R(\wn) = \pm it(\wn)$. Hence all the scattering coefficients are equal at $\wn$:
\begin{equation} \label{eq:EP_S_feature}
|r_L(\wn)|^2 = |r_R(\wn)|^2 = |t(\wn)|^2.
\end{equation}
This signature of an EP of $S$ at zero can thus be observed simply with standard one-sided reflection and transmission measurements.
The scattering behavior of the structure shown in Fig.~1g-i shows precisely the triple degeneracy of the scattering coefficients characteristic of an EP of $S$ at zero (Eq.~\ref{eq:EP_S_feature}).
This is initially surprising, since its parameters were chosen to be at CPA EP, not at an EP of $S$.
The structure differs from that of Fig.~1e only by the imposition of identical Bragg end mirrors. 

To understand why for this structure CPA EP and an EP of $S$ coincide we use temporal-coupled mode theory (TCMT)~\cite{Haus:1984uu}, which provides an analytic but approximate relationship between the eigenfrequencies of the wave operator and the $S$-matrix.  
Within TCMT one can show (supplement S3) that when the two cavities have equal out-coupling rates, CPA EP {\it does} imply a simultaneous EP of the $S$-matrix; but not when the cavities have unequal out-coupling rates. 
Thus, essentially the same experimental set-up can test the properties of these two different types of absorbing EPs.
If the TCMT theory were exact, the two eigenvalues of $S$ would coincide precisely at $\wn$ and would not be analytic there, leading to a complicated, non-quartic behavior near CPA.  Due to the approximate nature of TCMT, we find a slight displacement of the EP of $S$ from CPA EP, not visible in the results of Fig.~1g.

Returning to generic CPA EP, we now explore higher dimensional structures, both in free space and guided wave geometries.
For the case of resonant EPs, there has been extensive study of perturbed and deformed disk resonators in 2D, for which the EP of whispering gallery modes (WGMs) directly implies a spatially chiral solution, corresponding to either clockwise (CW) or counterclockwise (CCW) circulations of waves in the disk~\cite{Wiersig:2011hs, Wiersig:2014bq, Cao:2015fv}.
These strongly chiral resonances have been probed experimentally through asymmetric backscattering and chiral laser emission~\cite{Peng:2016dd}.
We now show that CPA EP in such a system will lead to chiral absorption: perfect absorption for, e.g. CCW input, and substantial backscattering for CW.
We note that standard CPA in disk and sphere resonators has been studied previously~\cite{Baranov:2017jv, Noh:2012wx}.

\begin{figure}[t]
\centering
\includegraphics[width=\linewidth]{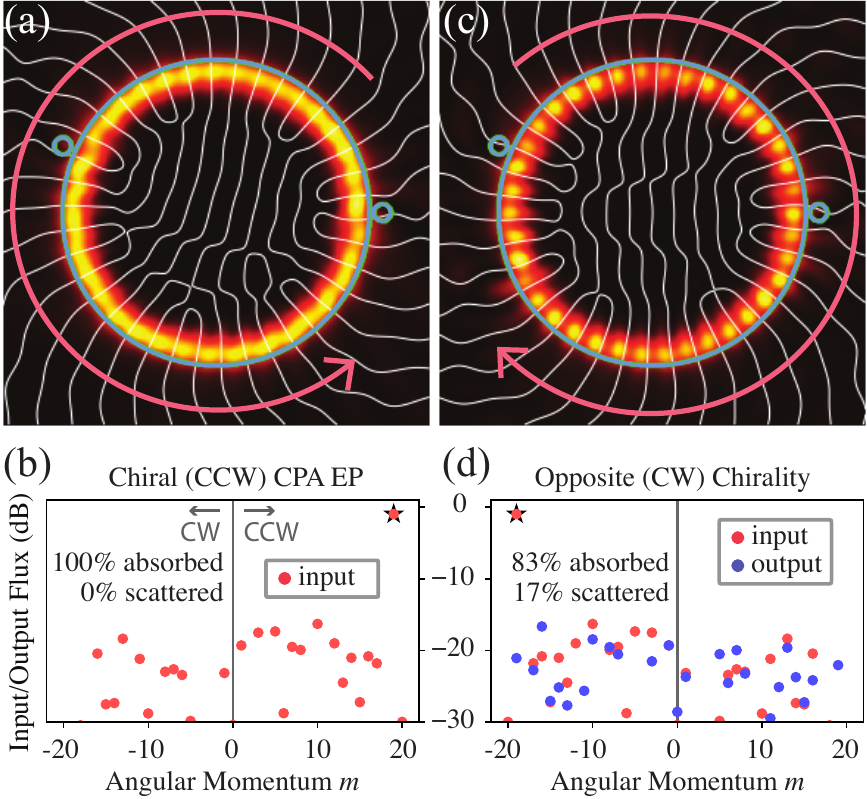}
\caption{
(Color online) Chiral CPA EP of WGMs of absorbing microdisk perturbed by point scatterers.
\textbf{(a)} CCW incident CPA EP mode.
Intensity plotted as color scale, curves of constant phase (white), disk boundary and scatterers (blue).
Curvature of phase fronts shows sense of rotation, denoted by arrow.
Uniform intensity along rim indicates running wave in disk.
\textbf{(b)} Input fluxes carried in each angular momentum channel for CPA EP (CCW) input.
Dominant channel (denoted by star) carries $80$\% of flux; CPA input is $>99.9$\% absorbed.
\textbf{(c)} Total field (incident \& scattered) for reverse chirality input.
Internal intensity shows standing wave oscillations due to presence of backscattering.
\textbf{(d)} As in (b), input fluxes (red), output (blue), for CW input.
Here we find $\sim 1$\% scattering across many channels, giving a total of $17$\% scattered flux ($83$\% absorption).
}
\label{fig:free space}
\end{figure}

We first consider an example of chiral absorption in free space, adapting the Wiersig model of a dielectric disk perturbed by two point scatterers~\cite{Wiersig:2011hs}, with parameters chosen to realize an absorbing EP at a real frequency (see supplement S1).
The perturbation from the first point scatterer splits the degenerate WGMs at angular momenta $m =\pm q$ into two standing-wave resonances, and fine-tuning the perturbation due to the second scatterer brings these two resonances back to degeneracy, forming an EP with CCW chirality at a complex frequency.
Finally, introducing a critical degree of absorption brings the absorbing EP to a real frequency.
As the scatterers break the rotational symmetry of the structure, the CPA EP input involves a
coherent superposition of many angular momenta other than $\pm q$, although at significantly weaker amplitude.
For the example shown in Fig.~2, the perfectly absorbed state has $80$\% of its incident flux at $q =19$, with the remaining $20$\% distributed across both CW and CCW at other $m$'s.
We test the chirality of absorption by exciting the disk with the corresponding CW input 
by exchanging $c_m \leftrightarrow (-)^m c_{-m}$ in the superposition; whereas the original state is $100\%$ absorbed, the opposite chirality is only $83\%$ absorbed.
Moreover, if we approximate the CPA input state by just its dominant component ($m=19$), both chiralities are equally absorbed (81\%).

The wavefront of the above free-space chiral CPA can be readily generated for acoustic waves, but an optical implementation may be challenging.
Therefore, we next consider chiral CPA EPs that are coupled in through a waveguide or fiber (see Fig.~3).
To reach CPA with a waveguide-only input, the free-space scattering loss rate should be much smaller than the waveguide coupling rate. Thus using point scatterers as tuning perturbations is undesirable, as they introduce additional scattering to free space.
Therefore instead of point scatterers, we introduce an azimuthally varying grating on the real and imaginary parts of the refractive index to promote the non-hermitian asymmetric coupling via absorption loss.
The system is well-modeled by TCMT, taking into account only the two single-mode running wave solutions in the fiber and the CW and CCW angular momentum states in the disk, coupled via the grating.
This configuration is similar to those used to study $\mathcal{PT}$-symmetry breaking and unidirectional invisibility in refs.~\cite{Lin:2011ij, Regensburger:2012jm, Feng:2013jj, Feng:2014gp, Feng:2014gg, Miao:2016eo}, but here we do not introduce any gain into the grating, only variable loss and a varying real part of the index, with no $\mathcal{PT}$-like discrete symmetries (see supplement S4B).

\begin{figure}[t]
\centering
\includegraphics[width=\linewidth]{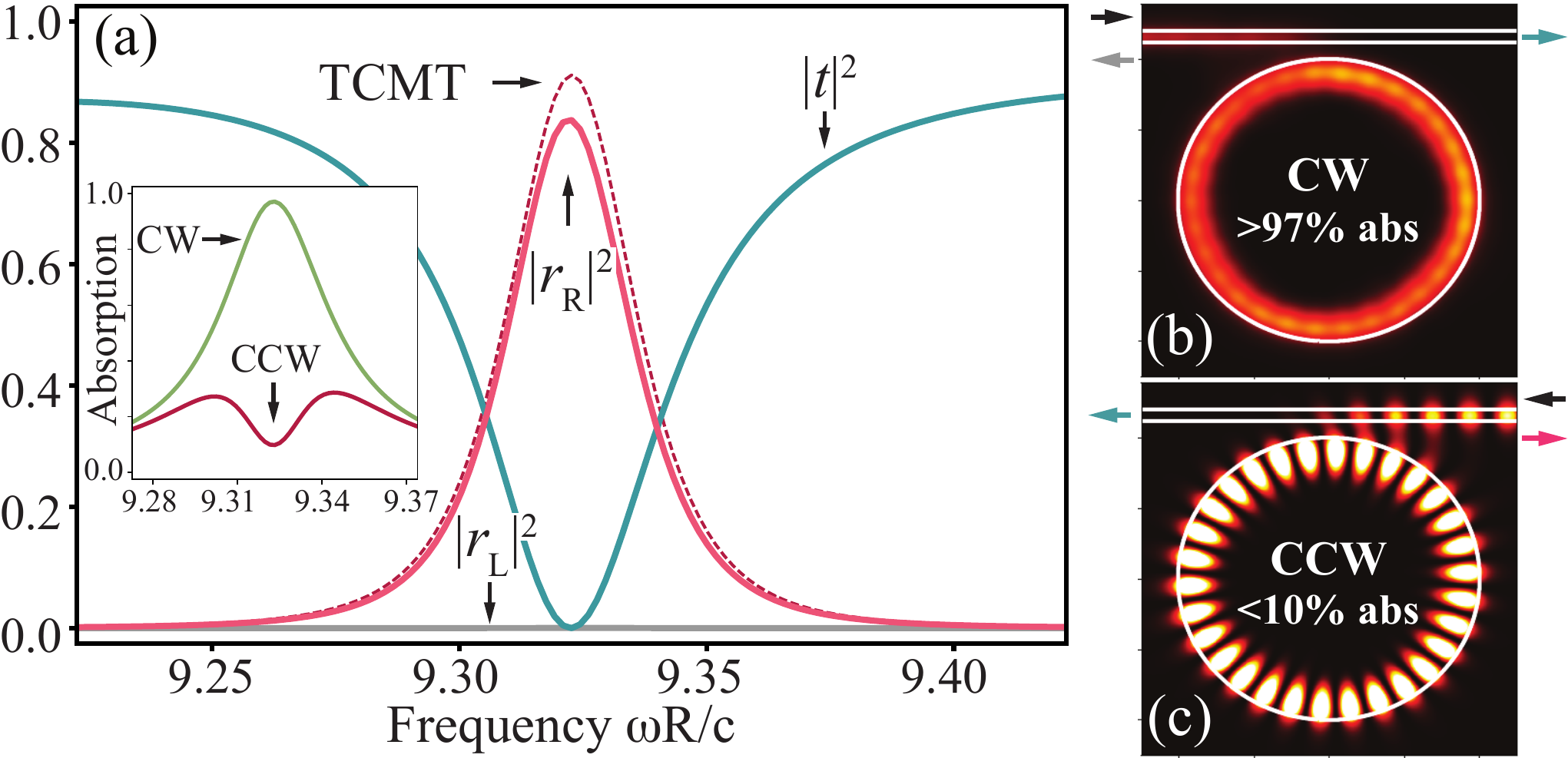}
\caption{
(Color online) Chiral absorption from CPA EP waveguide-microdisk system.
\textbf{(a)} Reflection $|r_L|^2$ (gray), $|r_R|^2$ (red solid), and transmission $|t|^2$ (blue) for disk with complex index azimuthal grating, tuned to CPA EP.
TCMT prediction (Eq.~\ref{TCMT main result}) for $r_R$ in dashed red.
Inset: one-sided absorption spectrum for left illumination (green) and right (red).
\textbf{(b)} Intensity of total field (incident \& scattered) for left illumination, corresponding to CPA EP.
\textbf{(c)} Same as (b) but for right illumination. Note standing wave of (c) vs running wave of (b) indicates strong coupling between CW and CCW modes only for right incidence, causing chiral absorption.
}
\label{fig:waveguide}
\end{figure}

The dielectric grating in Fig.~3 has a separable form $\delta\varepsilon = \rho(r)\tau(\theta)\varepsilon_0$ and couples WGMs with angular momenta $m=\pm q$ via azimuthal Fourier components $\tau_{\pm 2q}$, where $\tau(\theta)=\sum_n\tau_n e^{in\theta}$ ($\varepsilon_0$ is the dielectric function of the disk without the grating).
To achieve EP, one of the $\pm 2q$ components must vanish while the other remains finite (see supplement S4A), which can only occur with a complex index grating.
With the grating choice in Fig.~3, $\tau_{-2q}=0$, implying that the right propagating (CW) input at CPA EP will be strongly absorbed with negligible reflection, while $\tau_{2q} \neq 0$ will cause partial reflection of the left propagating (CCW) input.
Experimentally relevant gratings are piecewise constant, and in the simplest case have real and imaginary parts with the same angular width $\phi$ and periodicity $2\pi/P$, and an angular offset $\chi$ between them.
In this case, we show in the supplementary material (S4B) that an EP for WGMs with $m=\pm q$ is achieved when the real and the imaginary gratings have the same modulation magnitude, offset $\chi=(M-1/4)\pi/q$, and where $P$ divides $2q$ ($M,P\in\mathbb{Z}$).
Critically coupling the waveguide to the disk yields the desired CPA EP, with $r_L(\delta)=0$ and 
\begin{equation} \label{TCMT main result}
t(\delta) = \frac{\delta}{\delta+i\Gamma}, \quad
|r_{\rm R}(\delta)|^2 = \left(\frac{\sin q\phi}{q\phi}\right)^{2}\frac{1}{(1+\delta^2/\Gamma^2)^2},
\end{equation}
where $\delta=\omega-\omega_0$ is the detuning from the CPA EP frequency, and $\Gamma$ is the HWHM of the dip in $|t^2|$.
Note that to maximize reflection from the right (and minimize absorption), thinner lines of the grating are better, as this allows the standing wave to align its nodes with the narrower absorbing regions.
The reflection lineshape is a squared Lorentzian ~\cite{Pick:2017em}, while the transmission lineshape remains Lorentzian (supplement S4A).  As expected, the eigenchannel of $S$ (which is two-sided except at $\wn$) exhibits a quartic lineshape (not shown).

If we turn now to the results of the exact finite-difference frequency-domain numerical calculations (Fig.~3), we see that indeed the absorption in this geometry is strongly chiral, being  $>97\%$ when the disk resonance is excited from the left (CW excitation), but $<10\%$ when it is excited from the right (CCW excitation), the difference appearing predominantly as backscattering into the waveguide as expected, and in good agreement with the TCMT model.  Note that the $2.7\%$ of the input which is not absorbed for the CW excitation is removed by free space radiation and the CW reflection is truly negligible (supplement S5).

\begin{acknowledgments}
\paragraph{Acknowledgements}
We gratefully acknowledge useful discussions with Lan Yang, Changqing Wang, Mengzhen Zheng, and Liang Jiang. A.D.S. acknowledges support under NSF Grant No. DMR-1743235.
\end{acknowledgments}


\bibliography{Bibliography.bib}

\end{document}


\title{Supplementary Material for Perfectly absorbing exceptional points and chiral absorbers}

\author{William R. Sweeney}
\email{william.sweeney@yale.edu}
\affiliation{Department of Physics, Yale University, New Haven, CT 06520, USA}
\author{Chia Wei Hsu}
\affiliation{Department of Applied Physics, Yale University, New Haven, CT 06520, USA}
\author{Stefan Rotter}
\affiliation{Institute for Theoretical Physics, Vienna University of Technology (TU Wien), A-1040 Vienna, Austria, EU}
\author{A. Douglas Stone}
\affiliation{Department of Applied Physics, Yale University, New Haven, CT 06520, USA}
\affiliation{Yale Quantum Institute, Yale University, New Haven, CT 06520, USA}

\maketitle

\section*{S1. EP Parameter Values}

Here we state the parameter values used for the calculations reported in each of the figures. Bold type indicates parameters that were used in tuning to the EP.

\begin{table}[h]
\caption{ \label{tab:table 1}
EP Parameters for Fig 1 (bold indicates fine-tuned)
}
\begin{ruledtabular}
\begin{tabular}{ l r r r}
\textrm{ }  &  \textrm{Fig.~a-c}  &  \textrm{Fig.~d-f}  & \textrm{Fig.~g-i}\\
\colrule
Grating high index & 2                     & 2                     & 2\\
Grating low index  & 1.5                  & 1.5                  & 1.5\\
$L_1$        & {\bf1.2500}      & {\bf1.2560}      & 1.2566\\
$L_2$        & 1.4                  & 1.2566             &1.2566\\
$n_1^\prime$        & 2 & 2 & {\bf1.9981}\\
$in_1^{\prime\prime}$        & {\bf0.0382i} & {\bf0.0043i} & {\bf0.0037i}\\
$n_2^\prime$        & 2 & 2 & 2\\
$in_2^{\prime\prime}$        & {\bf0.0192i} & {\bf0.0472i} & {\bf0.0609i}\\
EP frequency $\wn$        & 5.0199             & 5.0012            & 5.0022\\
\end{tabular}
\end{ruledtabular}
\end{table}

\begin{table}[h]
\caption{ \label{tab:table 2}
EP Parameters for Fig 2 (bold indicates fine-tuned)
}
\begin{ruledtabular}
\begin{tabular}{ l r}
Index of disk                      & 1.5+{\bf0.0021i}  \\
Radius of disk                    & 1                      \\
Index of scatterers             & 1.5                      \\
Radius of scatterers          & 0.05                  \\
Distance of scatter 1         & 0.04      \\
Distance of scatter 2         & {\bf0.0454}                  \\
Angle between scatterers & {\bf156.30$^\circ$} \\
EP frequency $\wn$                                & 15.126             \\
\end{tabular}
\end{ruledtabular}
\end{table}

\begin{table}[h]
\caption{ \label{tab:table 3}
EP Parameters for Fig 3 (bold indicates fine-tuned)
}
\begin{ruledtabular}
\begin{tabular}{ l r}
Disk index                       & 2  \\
WGM mode number $q$ & 15 \\
Grating real index high   & {\bf2.0149} \\
Grating imag index high  & {\bf0.0153}\\
Grating width $\phi$        & 2$^\circ$ \\
Offset angle $\chi$          & {\bf8.9400$^\circ$}    \\
Grating periodicity $P$    & $30=2q$    \\
Disk radius                      & 1                      \\
Waveguide width             & 0.08                  \\
Waveguide distance         & 0.16                  \\
EP frequency $\wn$                               & 9.3230             \\
\end{tabular}
\end{ruledtabular}
\end{table}

\section*{S2. Derivation of Integral Relation for Helmholtz EP}

In this section we derive a previously-noted \cite{YaZeldovich:1961wz, Lai:1990tn, Chang:1996vz} relation between $\hat s\cdot \hat s$, where $\hat s$ is an eigenvector of the $S$-matrix, and the overlap integral (loosely speaking, ``inner product") of the eigenfunctions of a wave operator.
For simplicity, we focus on the scalar Helmholtz operator in one-dimension, over a domain of length $2L$.

Begin with two (nearby) solutions of the wave equation $\psi_{1,2}$ with eigenvalues $\omega_{1,2}$ ($c=1$) with incoming boundary conditions (appropriate for CPA) $\nabla\psi_{1,2}(\pm L)=\mp i\omega_{1,2}\psi_{1,2}(\pm L)$.
Consider the integral
\begin{equation}
\tag{S1}
(\psi_2,\{\nabla^2+\omega_1^2\varepsilon(x)\}\psi_1)=0.
\end{equation}
By integrating by parts twice, applying the boundary conditions, and dividing by a common factor of $(\omega_2-\omega_1)$ we have
\begin{equation}
\tag{S2}
c_0\hat s_1\cdot \hat s_2 = -i(\omega_2+\omega_1)\int dx\,\psi_2\,\varepsilon\,\psi_1,
\end{equation}
where $c_0^{2}=[\psi_1^2(-L)+\psi_1^2(L)][\psi_2^2(-L)+\psi_2^2(L)]$, and $\hat s_{1,2}\propto(\psi_{1,2}(-L),\psi_{1,2}(L))$ are the normalized $S$-matrix eigenvectors at $\omega_{1,2}$ with eigenvalue equal to zero.

The dielectric function can be parametrically deformed to bring about an accidental degeneracy (EP), so that $\omega_2\rightarrow \omega_1\equiv \omega_0$, and $\psi_2\rightarrow \psi_1 \equiv \psi_0$, in which case
\begin{equation}
\tag{S3}
c_0 \hat s_0\cdot \hat s_0 = -2i\omega_0\int dx\,\psi_0\,\varepsilon\,\psi_0.
\end{equation}

\section*{S3. Coincidence of EPs of $S$ and $H$ in TCMT for symmetric outcoupling}

In TCMT, the $S$-matrix is related to an effective Hamiltonian $H$ (not necessarily hermitian) by
\begin{equation}
\tag{S4} \label{general S-matrix} S = [1-2iW^\dagger\frac{1}{\omega-(H-iWW^\dagger)}W]S_0,
\end{equation}
where $S_0$ is the ``background'' scattering matrix, i.e. $S$ in the absence of resonances, and $W_{ij}$ is a matrix of coupling coefficients between mode $i$ and asymptotic channel $j$.

In the case where there are as many modes as there are asymptotic channels, $W$ is square. Additionally, if each mode couples to exactly one distinct channel, and all outcoupling rates are identical, then $W=\sqrt{\gamma/2}\,\mathbb{I}$. Then equation~\eqref{general S-matrix} reduces to
\begin{equation} \tag{S5} \label{specific S-matrix}
SS_0^{-1} = 1-\frac{i\gamma}{(\omega+i\gamma/2)-H},
\end{equation}

Now we apply a perturbation which tunes $H$ to a non-hermitian degeneracy. Then we can generally write $H=\Omega_{\mathrm{EP}}+N$, where $\Omega_{\mathrm{EP}}$ is the perturbed frequency, still degenerate, and $N$ is nilpotent: $N\neq0$ but $N^2=0$. We can expand the denominator as a geometric series, which truncates at $N^2$:
\begin{equation} \tag{S6} \label{N truncation}
\left[ \mathbb{I}-\frac{N}{(\omega-\Omega_{\mathrm{EP}}+i\gamma/2)} \right]^{-1} = \mathbb{I}+\frac{N}{(\omega-\Omega_{\mathrm{EP}}+i\gamma/2)}
\end{equation}
so that
\begin{equation} \tag{S7} \label{specific S-matrix}
SS_0^{-1} = D-\frac{i\gamma}{(\omega-\Omega_{\mathrm{EP}}+i\gamma/2)^2}N,
\end{equation}
where $D$ is some diagonal matrix. This makes $SS_0^{-1}$ manifestly exceptional.

For one-dimensional structures, such as the structures in Fig.~1 in the main text, there is no non-resonant coupling of left and right channels, and $S_0\propto \mathbb{I}$, so that if $SS_0^{-1}$ has an EP, then so too must $S$. Therefore in the geometry of Fig.~1g-i, with symmetric outcoupling, an EP of the wave operator (in the TCMT approximation this means an EP of $H$) implies a simultaneous EP of $S$.

\section*{S4. Calculation of Scattering Amplitudes at CPA EP} \label{supplemental Reflection}

\subsection*{A. TCMT for azimuthal perturbation}

In this section we derive the scattering coefficients for the waveguide-coupled microdisk at CPA EP using the coupled-mode framework.

First we consider a pair of degenerate modes of the unperturbed disk, clockwise (CW) and counterclockwise (CCW), which have angular momentum quantum numbers $-q$ and $q$, respectively. The degenerate complex frequency of the modes is $\Omega_0$. Additionally, each mode couples to one asymptotic channel of the waveguide with the same rate $\gamma$: CW to the right channel, and CCW to the left, so that in Eq.~\eqref{general S-matrix}, $W=\mathrm{diag}(\sqrt{\gamma/2},\sqrt{\gamma/2})$.
The waveguide is perfectly transmitting in the absence of the pair of resonances, so the non-resonant scattering matrix is
\begin{equation*}
S_0=
\begin{pmatrix} 
  0 & 1 \\ 
  1 & 0
\end{pmatrix}.
\end{equation*}

Upon right-multiplying both sides of Eq.~\eqref{general S-matrix} by $S_0$, we get a relation for the scattering amplitudes:
\begin{equation}
\tag{S8}
\label{SS0}
\begin{pmatrix}
 t & r_\mathrm{L} \\
 r_\mathrm{R} & t
\end{pmatrix}=1-\frac{i\gamma}{\omega-(H-i\gamma/2)}=\frac{\omega-(H+i\gamma/2)}{\omega-(H-i\gamma/2)}.
\end{equation}
Since we have not yet specified $H$, this applies to both the disk with and without the grating perturbation, though with different Hamiltonians.

If we bring $H$ to an EP by tuning parameters, we can apply Eq.~\eqref{N truncation}
so that
\begin{equation}
\label{lineshapes}
\tag{S9}
\begin{pmatrix}
 t & r_\mathrm{L} \\
 r_\mathrm{R} & t
\end{pmatrix} = \frac{\delta-i(\gamma-\Gamma)/2}{\delta+i(\gamma+\Gamma)/2}
-\frac{i\gamma}{(\delta+i(\gamma+\Gamma)/2)^2} N
\end{equation}
where $\delta$ is the detuning, and $\Gamma$ the overall loss rate:
\begin{equation}
\nonumber \delta=\omega-\mathrm{Re}\{\Omega_{\mathrm{EP}}\}  \qquad\quad  \Gamma/2=-\mathrm{Im}\{\Omega_{\mathrm{EP}}\}.
\end{equation}
Eq.~\eqref{lineshapes} fully characterizes the reflection and transmission coefficients as functions of frequency near a CPA EP, in terms of the nilpotent matrix $N$.

We now turn our attention to the calculation of this matrix in terms of a perturbation applied to the microdisk.
Under a perturbation $V$ the eigenvalues of a degenerate effective Hamiltonian $H$ shift by $\delta \Omega$: $H\rightarrow H+V$, $\Omega_0 \rightarrow \Omega_0+\delta \Omega$.
On the other hand, when the wave operator $\hat A=-\varepsilon^{-1}\nabla^2$ is perturbed by $\delta \hat A$, its spectrum shifts as $\Omega_0^2 \rightarrow (\Omega_0 + \delta \Omega)^2 \simeq \Omega_0^2 +2\Omega_0 \delta \Omega$.
It follows that small perturbations in the effective Hamiltonian and the wave operator are related by $V=\delta \hat A /2\Omega_0$.

For the case of the microdisk, we will first limit ourselves to separable perturbations $\varepsilon\rightarrow[1+\rho(r)\tau(\theta)]\varepsilon$, for which $\delta \hat A=-\rho(r)\tau(\theta) \hat A$.
The perturbation $V$, in the basis of the unperturbed Hamiltonian, and using the original eigenvalue equation $\hat A \psi = -\Omega_0^2 \psi$, is
\begin{equation} \tag{S10}
\label{Vmn}
V_{mn}=\frac{\Omega_0}{2}\int d^2 x\ \phi_m(\bx) \rho(r)\tau(\theta)\psi_n(\bx).
\end{equation}

The operator $\hat A$ is symmetric, therefore the sets of left and right eigenfunctions ($\{\phi\},\,\{\psi\}$, respectively) are equal and biorthogonal with weight $\varepsilon$, i.e. $\int d^2 x\ \phi_i\,\varepsilon\, \psi_j \propto \delta_{ij}$, usually written $(\phi_j,\psi_i)=\delta_{ij}$.
The eigenfunctions of $\hat A$ for the unperturbed microdisk are $\psi_m(r,\theta) = R_m(r)\exp(im\theta)$, and so by biorthogonality $\phi_m = \psi_{-m}$.
The matrix elements given by Eq.~\eqref{Vmn} can be evaluated in terms of the Fourier components of $\tau(\theta)$:
\begin{equation} \tag{S11} \label{Vmn2}
V_{mn}= \Omega_0 C_{mn}\tau_{m-n},
\end{equation}
where $C_{mn}=\pi\int_0^\infty dr\,r\, R_m R_{-n} \rho$ and $\tau(\theta)=\sum_n\tau_n e^{in\theta}$.
The effective Hamiltonian of the perturbed disk, in the degenerate CW/CCW basis, is therefore
\begin{equation} \tag{S12} \label{effective hamiltonian}
H_{mn} = \Omega_0 \delta_{mn}+ \Omega_0 C_{mn} \tau_{m-n}.
\end{equation}
To relate this to $S$ in Eq.~\eqref{lineshapes}, we make the assignment $\Omega_{\mathrm{EP}} = \Omega_0(1 + C \tau_0)$, and $N_{mn}=(1-\delta_{mn})\Omega_0 C_{mn} \tau_{m-n}$.
For the disk, the radial functions $R_m$ are given by Bessel functions of integer order, so that $R_m$ and $R_{-m}$ are related by a phase factor, and therefore so too are $C_{m,m}$ and $C_{-m,-m}$.
Therefore the $C$'s cannot be used to make $N^2=0$ with $N\neq0$.
To achieve this, it is instead necessary that exactly one of $\tau_{\pm2q}=0$.
This requires that $\tau(\theta)\notin \mathbb{R}$, otherwise $\tau_m=\tau_{-m}^*$ and both $\tau$'s would vanish.
This is where non-hermiticity is important for EP.
Without loss of generality, take $\tau_{-2q}=0$, so that $N_{q,-q}=\tau_{2q}\neq0$, with all other elements of $N$ vanishing.
Plugging this into Eq.~\eqref{lineshapes} and requiring CPA ($\gamma=\Gamma$), we determine the lineshapes of the reflection and transmission coefficients at CPA EP:
\begin{equation*}
t(\delta) = \frac{\delta}{\delta+i\Gamma},  \qquad   r_\mathrm{L}(\delta) = 0
\end{equation*}
\begin{equation} \tag{S13} \label{r(delta)}
r_\mathrm{R}(\delta) = \frac{i}{\Gamma}\frac{\Omega_0 C_{qq}\tau_{2q}}{(1-i\delta/\Gamma)^2} = \frac{r(0)}{(1-i\delta/\Gamma)^2}
\end{equation}
where $\Gamma=-2\mathrm{Im}\{\Omega_0(1+C_{qq}\tau_0)\}$.
The amplitudes for transmission and reflection in the ``correct'' direction vanish exactly as they would for CPA or critical coupling in the absence of an EP.
The remaining reflection amplitude for the ``wrong" direction of incidence, at the CPA EP frequency ($\delta=0$), is
\begin{equation} \tag{S14} \label{rd}
r_\mathrm{R}(0) = - \frac{i}{2}\frac{C_{qq}\tau_{2q}(2+iQ_0^{-1})}{2\mathrm{Im}\{C_{qq}\tau_0\}+Q_0^{-1}(1+\mathrm{Re}\{C_{qq}\tau_0 \})}.
\end{equation}
$Q_0$ is the quality factor of the bare disk, without grating or waveguide: $Q_0=-\mathrm{Re}\{\Omega_0\}/2\mathrm{Im}\{\Omega_0\}$.
In the limit where the bare disk resonances have $Q_0\gg1$ (which is typical for WGMs), we can neglect the $Q_0^{-1}$ terms. In this limit we also approximate the radial integral $C_{qq}$ to be real.
Hence the nontrival reflection amplitude in the high-$Q$ limit takes the remarkably simple form
\begin{equation} \tag{S15} \label{r0}
|r_\mathrm{R}(0)|^2 = \frac{1}{4}\frac{|\tau_{2q}|^2}{\mathrm{Im}\{\tau_0\}^2}.
\end{equation}
The overall gain/loss added to the system is encoded in $\tau_0$, which is therefore determined by the critical coupling condition.

The analysis can be extended to include non-separable perturbations, so long as they can be decomposed into separable pieces: $\delta\varepsilon(r,\theta) = \varepsilon \sum_j \rho^j(r)\tau^j(\theta)$. The nilpotent matrix becomes $N_{mn}=(1-\delta_{mn})\Omega_0 \sum_j C^j_{mn}\tau^j_{m-n}$, where $C^j_{mn}=\pi\int_0^\infty dr\,r\, R_m R_{-n} \rho^j$ and $\tau^j(\theta)=\sum_n\tau^j_n e^{in\theta}$. The condition for $N$ nilpotent is that only one of $N_{\pm q,\mp q}$ vanish, say $N_{-q,q}$: $\sum_j C^j_{-q,q}\tau^j_{-2q}=0$, but $\sum_j C^j_{q,-q}\tau^j_{2q}\neq0$. In this case we no longer need a non-hermitian perturbation to achieve EP, though we must rely on the radial integrals ($C$'s) being complex. The point scatterers used in Fig.~2 exemplify this: a purely real, (approximately)~separable set of perturbations that support EP.

\subsection*{B. Engineering for maximal asymmetry of reflection and absorption}

It is evident from Eq.~\eqref{r0} that scattering from the waveguide-disk system is entirely characterized by the two Fourier components of the perturbation $\tau_0$ and $\tau_{2q}$, which suggests that the appropriate design to consider is a non-hermitian grating.
We consider only gratings with no gain, with alternating regions of loss and no loss.
The simplest experimentally feasible azimuthal grating of this type is piecewise constant, whose real and imaginary parts have the same angular width $\phi$ and periodicity $2\pi/P$ ($P\in\mathbb{Z}$), and an angular offset $\chi$ between them:
\begin{equation} \tag{S16} \label{tau}
\tau(\theta)=f(\theta)+if(\theta-\chi),
\end{equation}
where $f(0<\theta<\phi) = c$, $f(\phi<\theta<2\pi/P)=0$, and $f(\theta+2\pi/P)=f(\theta)$, for some constant $c$.
The angular offset $\chi$ is determined from $\tau_{-2q}=(1+ie^{2iq\chi})f_{2q}=0$, which implies
\begin{equation} \tag{S17} \label{chi}
\chi = (M-1/4)\pi/q.
\end{equation}
Of course had we demanded the $+2q$ component to vanish, this would be $\chi=(M+1/4)\pi/q$.

We can express the reflection from the ``incorrect'' side in terms of $f_m$ according to Eq.~\eqref{r0}, and using Eq.~\eqref{chi}:
\begin{equation} \tag{S18} \label{newr0}
r_\mathrm{R}(0) = \left| \frac{f_{2q}}{f_0} \right|^2
\end{equation}
The Fourier components of $f$ vanish for $m$ not equal to a multiple of $P$; the non-vanishing components satisfy
\begin{equation*} \label{fm}
f_{n\cdot P} = c\frac{P}{2\pi}\int_0^\phi d\theta e^{-inP\theta} = e^{-inP\phi/2}\frac{c}{n\pi} \sin \frac{nP\phi}{2}.
\end{equation*}
Plugging this into Eq.~\eqref{newr0} gives
\begin{equation} \tag{S19}
r_\mathrm{R}(0) = \left| \frac{ \sin q\phi}{q\phi} \right|^2,
\end{equation}
so long as $NP=2q$, where $N$ is the order of the grating that we are using to couple the $\pm q$ modes.

We see that the asymmetry of the reflection, and therefore of the absorption, achieves its maximal value, unity, for thin gratings ($\phi\rightarrow0$). The intuition is that the lossy regions can be ``hidden'' in the nodes of the back-scattered field when excited from the non-CPA side, and the thinner they are, the better they are hidden. Since the field is a running wave when excited from the CPA side, the material loss is just as effective regardless of how narrow its spatial distribution. This is evident in Fig.~3.

A more general type of grating has different widths, contrasts, and periodicities for its real and imaginary parts. If the real part has contrast $a$, periodicity $L$, width $\phi$, while the imaginary part has ($b$, $P$, $\psi$), the conditions for CPA EP are
\begin{equation*}
\left| \frac{\sin q\phi}{\sin q\psi} \right|=\frac{b}{a}\frac{P}{L},
\end{equation*}
$L$ and $P$ must both divide $2q$, and the offset is
\begin{equation*}
\chi = (M-1/4)\pi/q + (\phi-\psi)/2.
\end{equation*}
In this case, the asymmetric reflection is
\begin{equation*}
r_\mathrm{R}(0) = \left| \frac{ \sin q\phi}{q\phi} \right|^2 \cos^2(q[\phi-\psi]),
\end{equation*}
which shows that the more restrictive grating analyzed earlier ($\phi=\psi$) is optimal.

It is worth noting that gauged $\mathcal{PT}$-symmetry corresponds to $\phi=\pi/P$, which yields $|r_\mathrm{R}(0)|^2=\mathrm{sinc}^2(N\pi/2)<41\%$.

\section*{S5. Free Space Loss and Chiral CPA EP} \label{supplemental CPA EP}

The disk plus waveguide does not admit CPA solutions which only propagate in on the waveguide; 
the exact CPA solutions will require some small flux to excite the disk from free space, just as the corresponding laser would radiate weakly into those free space channels.  If we simply take the system at the CPA EP solution parameters but excite solely through the waveguide, we then do not expect to find 100\% absorption or zero transmission along the fiber; and indeed when we implemented this procedure we found a small, but measurable transmission.  Since we are interested in a chiral absorber without free-space excitation we hence adjusted the waveguide parameters in order to minimize this transmission, moving away from the exact CPA EP point.

Qualitatively we expect free-space channels to act as a small additional loss with respect to the guided channels. Therefore we increased the coupling to the fiber by a few percent until the transmission in the fiber was minimized, while otherwise maintaining the same structure, and found that the transmission became negligible.  Since we are no longer solving the exact CPA problem
we are no longer guaranteed that all the flux will be absorbed in the disk with grating, some will be lost to free space radiation. 
This is the reason why in Fig.~3 the absorption from the CPA EP is not unity, but it is still greater than 97\%.


\bibliography{Bibliography.bib}